\documentclass[12pt]{article}

\usepackage{fullpage}
\usepackage{graphics}
\usepackage{amssymb}
\newtheorem{theorem}{Theorem}[section]

\newtheorem{lemma}[theorem]{Lemma}

\newtheorem{corollary}[theorem]{Corollary}

\newtheorem{proposition}[theorem]{Proposition}

\newtheorem{definition}[theorem]{Definition}

\newtheorem{remark}[theorem]{Remark}

\newcommand {\floor}[1]{\ensuremath{\lfloor}{#1}\ensuremath{\rfloor}}
\newcommand {\ket} [1] {\ensuremath{\vert}{#1}\ensuremath{\rangle}}
\newcommand {\bra} [1] {\ensuremath{\langle}{#1}\ensuremath{\vert}}
\newcommand {\braket} [2]
{\ensuremath{\langle}{#1}\ensuremath\,{\vert}\,{#2}\ensuremath{\rangle}}

\newcommand{\F}{\ensuremath{\mathbb{F}}}
\newcommand{\BAD}{\ensuremath{\rm BAD}}

\newcommand{\bproof}{\noindent{\it Proof}}
\newcommand{\cproof}{\noindent{\it Proof of Claim}}
\newcommand{\eproof}{\hspace*{\fill}$\rule{2mm}{2mm}$~~~~~\bigskip}
\renewenvironment{proof}{\bproof. }{\eproof}

\newcommand{\Tr}{\mbox{\it Tr}}

\renewcommand{\bold}[1]{{\bf #1}}

\renewcommand{\a}{{\bf a}}
\renewcommand{\b}{{\bf b}}
\renewcommand{\c}{{\bf c}}
\renewcommand{\d}{{\bf d}}
\newcommand{\e}{{\bf e}}
\newcommand{\x}{{\bf x}}
\renewcommand{\v}{{\bf v}}
\newcommand{\y}{{\bf y}}
\renewcommand{\r}{{\bf r}}
\newcommand{\om}{{\omega}}
\newcommand{\w}{\tilde{\omega}}

\newcommand{\balpha}{\ensuremath{\mbox{\boldmath $\alpha$}}}
\newcommand{\bbeta}{\ensuremath{\mbox{\boldmath $\beta$}}}
\newcommand{\bchi}{\ensuremath{\mbox{\boldmath $\chi$}}}
\newcommand{\sbalpha}{\ensuremath{\mbox{\scriptsize\boldmath $\alpha$}}}
\newcommand{\sbbeta}{\ensuremath{\mbox{\scriptsize\boldmath $\beta$}}}
\newcommand{\sbchi}{\ensuremath{\mbox{\scriptsize\boldmath $\chi$}}}

\newcommand{\brac}[2]{{\ensuremath{[[{#1}]]_{{#2}}}}}

\newcommand{\wt}{{\it wt}}

\newcommand{\Hi}{{\ensuremath{\mathcal{H}}}}
\newcommand{\A}{{\ensuremath{\mathcal{A}}}}
\newcommand{\C}{{\ensuremath{\mathcal{C}}}}
\newcommand{\V}{{\ensuremath{\mathcal{V}}}}
\renewcommand{\S}{{\ensuremath{\mathcal{S}}}}
\newcommand{\Hin}{{\ensuremath{\mathcal{H}^{\otimes^n}}}}
\newcommand{\HinA}{{\ensuremath{{L^2(A)}^{\otimes^n}}}}
\newcommand{\E}{{\ensuremath{\mathcal{E}}}}

\renewcommand{\th}[1]{\ensuremath{#1^{th}}}

\title{{\bf A Family of Quantum Stabilizer Codes Based on the Weyl
    Commutation Relations over a Finite Field}}

\author{
V.~Arvind and K.~R.~Parthasarathy\\
Institute of Mathematical Sciences, C.I.T Campus\\ 
Chennai 600113, India\\
email: {\tt\{arvind,krp\}@imsc.ernet.in}\\}

\date{}

\begin{document}

\maketitle
\vspace{-1cm}
\begin{center}
{\small\it Dedicated to C.S. Seshadri on his 70th birthday}
\end{center}

\begin{abstract}
  Using the Weyl commutation relations over a finite field $\F_q$ we
  introduce a family of error-correcting quantum stabilizer codes
  based on a class of symmetric matrices over $\F_q$ satisfying
  certain natural conditions. When $q=2$ the existence of a rich class
  of such symmetric matrices is demonstrated by a simple probabilistic
  argument depending on the Chernoff bound for i.i.d symmetric
  Bernoulli trials. If, in addition, these symmetric matrices are
  assumed to be circulant it is possible to obtain concrete examples
  by a computer program. The quantum codes thus obtained admit elegant
  encoding circuits.
\end{abstract}

\section{Introduction}

Let $A$ be a finite abelian group with operation denoted by $+$ and
identity 0. We identify $A$ with the alphabet of symbols transmitted
on a classical communication channel. Consider the $n$-fold cartesian
product $A^n$ of copies of $A$. Elements of $A^n$ are called words of
length $n$. A commonly used group is $\{0,1\}$ with addition modulo 2.
Let $\hat{A}$ denote the character group of $A$, the multiplicative
group of all homomorphisms {from} $A$ into the multiplicative group of
complex numbers of modulus unity. For $\a=(a_1,a_2,\ldots,a_n)^T\in A^n$
we define its weight $w(\bold{a})$ to be $\#\{i\mid a_i\neq 0\}$. We
say that a subgroup $\C_n$ of $A^n$ is a \emph{$t$-error correcting
  group code} if for every non-zero element
$\bold{x}=(x_1,x_2,\ldots,x_n)^T$ in $\C_n$, $w(\bold{x})\geq 2t+1$. In
other words, if messages transmitted through a noisy channel are
encoded into words {from} $\C_n$ and during transmission of a word
errors at the output occur in at most $t$ positions, then the message
can be decoded without any error. There is a vast literature on the
construction of $t$-error correcting group codes and the reader may
find an introduction to this subject and pointers to literature in
\cite{slone, vLin}.

A broad class of quantum error correcting codes known as stabilizer
codes was introduced by Gottesman \cite{gottes} and Calderbank et al
\cite{cald-shor} (also see \cite{kl2,nonadd,rains}). To the best of
our knowledge, apart {from} one computer-generated example
\cite{nonadd}, all quantum error-correcting codes are stabilizer
codes. Our aim is to give a new description of the theory of
error-correcting quantum stabilizer codes. First we introduce some
definitions. We choose and fix an $N$-dimensional complex Hilbert
space $\Hi$ and consider the unit vectors of $\Hi$ as pure states of a
finite level quantum system. If $A$ is a finite abelian group with $N$
elements and $\{e_x\mid x\in A\}$ is an orthonormal basis of $\Hi$
indexed by elements of $A$ we express it in the Dirac notation as
$\ket{x}=e_x$.  If $\bold{x}=(x_1,x_2,\ldots,x_n)^T\in A^n$ is a word
of length $n$, we write
\[
\ket{\x}=\ket{x_1x_2\ldots x_n}=e_{x_1}\otimes e_{x_2}\otimes\ldots\otimes e_{x_n}
\]
where the right-hand side is a product vector in the $n$-fold tensor
product $\Hi^{\otimes^n}$ of $n$ copies of $\Hi$. Thus, with the
chosen orthonormal basis, every word $\x$ in $A^n$ is translated into
a basis state $\ket{\x}$ of $\Hi^{\otimes^n}$.

A \emph{quantum code} is a subspace $\C_n$ in $\Hi^{\otimes^n}$. Note
that a pure state in $\Hi^{\otimes^n}$ described by a unit vector
$\ket{\psi}$ in $\Hi^{\otimes^n}$ has density matrix
$\ket{\psi}\bra{\psi}$. A density matrix $\rho$ in $\Hi^{\otimes^n}$
is a non-negative operator of unit trace. In quantum probability, a
projection operator $E$ in $\Hi^{\otimes^n}$ is interpreted as an
event concerning the quantum system and a density matrix $\rho$ as a
state of the quantum system. The probability of the event $E$ in the
state $\rho$ is given by $\Tr\rho E$. Messages to be transmitted
through a quantum channel are encoded into pure states in
$\Hi^{\otimes^n}$. When a pure state $\ket{\psi}$, or equivalently, a
density matrix $\ket{\psi}\bra{\psi}$ is transmitted the channel
output is hypothesized to be a state of the form
\begin{eqnarray}
\rho=\sum_{i}L_i\ket{\psi}\bra{\psi}L_i^{\dagger}\label{eq-1}
\end{eqnarray}
where the operators $\{L_i\}$ belong to a linear subspace $\A$ of the
algebra of all operators on $\Hi^{\otimes^n}$. The operators $\{L_i\}$
may depend on $\rho$, but in order to ensure that $\rho$ is a density
matrix it is assumed that $\bra{\psi}\sum_i
L^{\dagger}_iL_i\ket{\psi}=1$. By the spectral theorem $\rho$ can
be expressed as
\[
\rho = \sum_{j}p_j\ket{\psi_j}\bra{\psi_j}
\]
where  $\psi_j$  is  an   orthonormal  set  in  $\Hi^{\otimes^n}$  and
$\{p_j\}$ is a probability distribution with $p_j>0$ for each $j$.  In
other  words, the  output state  $\rho$ is  not necessarily  pure even
though  the  input state  is  pure.  The  operators $L_i$  are  called
\emph{error  operators} and the  linear space  $\A$ {from}  which they
come is called the \emph{ error space}.

Suppose there is a finite family $\{M_j\}$ of operators in
$\Hi^{\otimes^n}$ satisfying the condition $\sum_j M^{\dagger}_jM_j=I$
and for any output state $\rho$ with $\psi$ in the code $\C_n$,
\[
\sum_j M_j\rho M^{\dagger}_j=
\sum_{i,j}M_jL_i\ket{\psi}\bra{\psi}L_i^{\dagger}M_j{\dagger}
=\ket{\psi}\bra{\psi}.
\]
Then we say that the quantum code $\C_n$ together with the family
$\{M_j\}$ of 'decoding operators' corrects any error induced by
$\{L_i\}$ {from} $\A$. In this context we have the following fundamental
theorem of Knill and Laflamme \cite{KL} which gives necessary and
sufficient conditions for the existence of such a family of decoding
operators.

\begin{theorem}{\rm\cite{KL}}\label{kl-theorem}
  Let $\A$ be a family of operators in $\Hin$ and let $\C_n\subset
  \Hin$ be a quantum code with an orthonormal basis
  $\psi_1,\psi_2,\ldots,\psi_d$. Then there exists a finite family 
$\{M_j\}$ of operators in $\Hin$ satisfying the conditions:
\begin{enumerate}
\item[(i)] $\sum_j M^{\dagger}_jM_j=I$; and
\item[(ii)] 
\[\sum_{j}M_jL\ket{\psi}\bra{\psi}L^{\dagger}M_j^{\dagger}
  =\bra{\psi}L^{\dagger}L\ket{\psi}\ket{\psi}\bra{\psi}\mbox{~~}
  \forall\mbox{~~} \psi\in\C_n, L\in\A
\]
\end{enumerate}
if and only if the following condition holds:\\
$\bra{\psi_p}L_1^{\dagger}L_2\ket{\psi_q}=\delta_{p,q}c(L_1,L_2)$ for
all $L_1,L_2\in\A$, $1\leq p,q\leq d$, where $c(L_1,L_2)$ is a scalar
independent of $p$ and $q$ and $\delta_{p,q}$ is 1 if $p=q$ and 0
otherwise.
\end{theorem}

\begin{remark}
The proof of the above theorem is constructive and therefore yields
the decoding operators in terms of $\A$ and the basis
$\psi_1,\ldots,\psi_d$ of $\C_n$. In this case we say that $\C_n$ is
an $\A$-error correcting quantum code.
\end{remark}

Now we specialize the choice of $\A$. Consider all unitary operators
in $\Hin$ of the form $U=U_1\otimes U_2\otimes\ldots\otimes U_n$ where
each $U_i$ is a unitary operator on $\Hi$ and all but $t$ of the
$U_i$'s are equal to $I$. Such a $U$ when operating on
$\psi=\psi_1\otimes\ldots\otimes \psi_n\in\Hin$ produces $U\ket{\psi}$
which is an $n$-fold tensor product that differs {from} $\psi$ in at
most $t$ places. Denote by $\A_t$ the linear span of all such unitary
operators $U$. A quantum code $\C_n$ is called a \emph{$t$-error
  correcting quantum code} if $\C_n$ is an $\A_t$-correcting quantum
code. 

\section{Quantum codes and subgroups of the error group}\label{defs}

Let $(A,+)$ be a finite abelian group with $N$ elements and identity
denoted by 0. Denote by $\hat{A}$ the character group of $A$ and $\Hi$
the $N$-dimensional Hilbert space $L^2(A)$ of all complex-valued
functions on $A$, spanned by $\{\ket{x}\}_{x\in A}$ (where the vector
$\ket{x}$ denotes the indicator function $1_x$ of the singleton
$\{x\}$).  Define the unitary operators $U_{a}$ and $V_{\chi}$ on
$\Hi$ for every $a\in A$ and $\chi\in\hat{A}$ by
\[
U_a\ket{x}=\ket{x+a},\hspace{2cm} V_{\chi}\ket{x}=\chi(x)\ket{x}
\]
where $x\in A$. Then 

\[
\chi(a)U_aV_{\chi}=V_{\chi}U_a\mbox{~~}\forall\mbox{~~} a\in A, \chi\in\hat{A}.
\]
These are the Weyl commutation relations between the unitary operators
representing $A$ by translations and $\hat{A}$ by multiplications.
The family of operators $\{U_aV_{\chi}\mid a\in A \chi\in\hat{A}\}$ is
irreducible.

If $\a\in A^n$ then any element $\bchi\in\hat{A^n}$ can be identified with
an element of $\hat{A}^n$ so that
\[
\bchi(\a)=\prod_{i=1}^n\chi_i(a_i)~~~~~~~\chi_i\in\hat{A}, a_i\in A
\]
where $\bchi=(\chi_1,\ldots,\chi_n)$ and $\a=(a_1,\ldots,a_n)$. Put
$U_{\a}=U_{a_1}\otimes\ldots\otimes U_{a_n}$ and
$V_{\sbchi}=V_{\chi_1}\otimes\ldots\otimes V_{\chi_n}$. Then
$\{U_{\a}V_{\sbchi}\mid \a\in A^n, \bchi\in\hat{A}^n\}$ is again
an irreducible family of unitary operators satisfying the Weyl commutation
relations

\[
\bchi(\a)U_{\a}V_{\sbchi}=V_{\sbchi}U_{\a}\mbox{~~}\forall\mbox{~~} \a\in A^n, 
\bchi\in\hat{A}^n.
\]

In the Hilbert space of all linear operators on $\Hi^{\otimes^n}$
equipped with the scalar product $\braket{X}{Y}=\Tr X^{\dagger}Y$ the
set $\{N^{-n/2}U_{\a}V_{\bchi}\mid \a\in A^n, \bchi\in\hat{A}^n\}$ is
an orthonormal basis. The weight $\wt(\a,\bchi)$ of a pair
$(\a,\bchi)\in A^n\times \hat{A}^n$ is defined to be $\#\{i\mid 1\leq
i\leq n, (a_i,\chi_i)\neq (0,1)\}$, where $\a=(a_1,a_2,\ldots,a_n)$
and $\bchi=(\chi_1,\ldots,\chi_n)$. The irreducibility of
$\{U_{\a}V_{\sbchi}\mid \a\in A^n, \bchi\in\hat{A}^n\}$ implies that
$\{U_{\a}V_{\sbchi}\mid \a\in A^n, \bchi\in\hat{A}^n, \wt(\a,\bchi)\leq
t\}$ spans $\A_t$. The Knill-Laflamme theorem for $\A_t$-correcting
quantum codes assumes the following form which can be readily derived
from Theorem~\ref{kl-theorem}.

\begin{theorem}\label{kl-2}
  $\C_n\subset L^2(A)^{\otimes^n}$ is a $t$-error correcting quantum
  code if and only if $\C_n$ has an orthonormal basis
  $\psi_1,\psi_2,\ldots,\psi_d$ satisfying the following
  conditions:\\
  For every $(\a,\bchi)\in A^n\times\hat{A}^n$ such that
  $\wt(\a,\bchi)\leq 2t$
\begin{enumerate}
\item[(i)] $\bra{\psi_i}U_{\a}V_{\sbchi}\ket{\psi_j}=0$ if $i\neq j$,
  and 
\item[(ii)] $\bra{\psi_i}U_{\a}V_{\sbchi}\ket{\psi_i}$ is a
  scalar independent of $\psi_i$ for $i=1,2,\ldots,d$.
\end{enumerate}
\end{theorem}

Let $l$ be the least positive integer such that $la=0$ for all $a\in
A$ and let $\omega=e^{\frac{2\pi i}{l}}$. We define the \emph{error
  group} as the following finite group of unitary operators in
$L^2(A)^{\otimes^n}$.
\[
\E=\{\om^iU_{\a}V_{\sbchi}\mid 0\leq i\leq l-1, \a\in A^n, \bchi\in \hat{A}^n\}.
\]
The group $\E$ has a natural action on the Hilbert space
$L^2(A)^{\otimes^n}$ defined by:

\[
U_{\a}\ket{\x}=\ket{\x+\a},\hspace{2cm} V_{\sbchi}\ket{\x}=\chi(\x)\ket{\x}.
\]
Subspaces of $L^2(A)^{\otimes^n}$ that are point-wise fixed by some
subgroup of the error group $\E$ are called \emph{stabilizer codes}.

Let $\S$ be a subgroup of $\E$. Denote by $\C(\S)$ the subspace of
$L^2(A)^{\otimes^n}$ that is point-wise stabilized by $\S$. More precisely,
\[
\C(\S)=\{\psi\in\HinA\mid U\psi=\psi\mbox{~~}\forall\mbox{~~} U\in \S\}.
\]

\begin{lemma}\label{stab-1}
  $\C(\S)\neq 0$ if and only if $\S$ is an abelian subgroup of $\E$ such
  that $\om^iI\not\in \S$ for $i\neq 0$. Furthermore, when $\C(\S)\neq 0$
  the dimension of $\C(\S)$ is $\#A^n/{\#\S}$.
\end{lemma}

\begin{proof}
  Suppose $\om^iI\in \S$ for some $i\neq 0$. For any $\psi\in\C(\S)$ we
  have $\om^iI\psi=\psi$ , which implies $\psi=0$. Hence $\C(\S)=0$.
  
  It follows {from} the Weyl commutation relations that two elements
  $\om^iU_{\a}V_{\sbalpha}$ and $\om^jU_{\b}V_{\sbbeta}$ in $\S$ commute if
  and only if $\balpha(\b)=\bbeta(\a)$. Now, let $\psi\in\C(\S)$. We
  have
\[
\psi=\om^iU_{\a}V_{\sbalpha}\om^jU_{\b}V_{\sbbeta}\psi = 
\om^jU_{\b}V_{\sbbeta}\om^iU_{\a}V_{\sbalpha}\psi.
\]

Applying the commutation relations we can see that the above equation
holds for a $\psi\neq 0$ if and only if $\balpha(\b)=\bbeta(\a)$.
Thus, $\C(\S)\neq 0$ if and only if $\S$ is abelian and $\om^iI\not\in
\S$ for $i\neq 0$.

Now, let $\S$ be an abelian subgroup of $\E$ such that $\om^iI\not\in \S$
for $i\neq 0$. Define the projection operator
\[
P = \frac{1}{\#\S}\sum_{U\in \S} U.
\]
Since $\Tr U_{\a}V_{\sbbeta}=0$ unless $(\a,\bbeta)=(0,1)$ it follows
that $\Tr(P)=\#A^n/{\#\S}$. It is easy to see that $P$ is the
projection onto $\C(\S)$. Thus, the dimension of $\C(\S)$ is
$\Tr(P)=\#A^n/{\#\S}$.  This completes the proof.
\end{proof}

Next, we state Theorem~\ref{kl-2} in a form that will give the criteria
for constructing $t$-error correcting quantum stabilizer codes.
Let $Z(\S)$ denote the centralizer of $\S$ in $\E$, i.e.,

\[
Z(\S) = \{U\in\E\mid UU'=U'U \mbox{~~}\forall\mbox{~~} U'\in \S\}.
\]

\begin{theorem}\label{kl-3}
  Let $\S$ be an abelian subgroup of the error group $\E$ such that
  $\om^iI$ is not in $\S$ for $i\neq 0$. Then $\C(\S)$ is a $t$-error
  correcting quantum code if $\wt(\a,\balpha)>2t$ for each
  $\om^iU_{\a}V_{\sbalpha}\in Z(\S)\setminus\S$.
\end{theorem}

\begin{proof} 
  Suppose $\wt(\a,\balpha)>2t$ for each $\om^iU_{\a}V_{\sbalpha}\in
  Z(\S)\setminus\S$. Now, by the previous lemma $\C(\S)$ is a subspace of $\HinA$
  of dimension $\#A^n/{\#\S}=d$. Let $\psi_1,\ldots,\psi_d$ be an
  orthonormal basis of $\C(\S)$. Consider a $(\a,\bchi)\in
  A^n\times\hat{A}^n$ with the property that $\wt(\a,\bchi)\leq 2t$.
  We check the Knill-Laflamme conditions (Theorem~\ref{kl-2}). There
  are two cases:
\begin{enumerate}
\item[(a)]
If $\om^iU_{\a}V_{\sbchi}\in \S$ for some $i\geq 0$ then
\[
\bra{\psi_j}\om^iU_{\a}V_{\sbchi}\ket{\psi_k}=\braket{\psi_j}{\psi_k}=\delta_{jk},
\mbox{~~}1\leq j, k\leq d.
\]
Thus, $\bra{\psi_j}U_{\a}V_{\sbchi}\ket{\psi_k}=\om^{-i}\delta_{jk}
1\leq j, k\leq d$, where $\delta_{jk}$ is the Kronecker delta
function.
\item[(b)] If $\om^iU_{\a}V_{\sbchi}\not\in \S$ for each $i\geq 0$, then
  since $\wt(\a,\bchi)\leq 2t$, $\om^iU_{\a}V_{\sbchi}\not\in Z(\S)$
  for each $i\geq 0$ by the assumption. Let $\psi\in\C(\S)$ and
  $\om^rU_{\b}V_{\sbbeta}$ be some element of $\S$. Then we can write
$\bra{\psi}U_{\a}V_{\sbchi}\ket{\psi}$ as
$\bra{\om^rU_{\b}V_{\sbbeta}\psi}U_{\a}V_{\sbchi}\ket{\om^rU_{\b}V_{\sbbeta}\psi}$,
which can be simplified to get the following
\begin{eqnarray}\label{eq-2}
\bra{\psi}U_{\a}V_{\sbchi}\ket{\psi}=
\overline{\bbeta(\a)}\bchi(\b)\bra{\psi}U_{\a}V_{\sbchi}\ket{\psi}.
\end{eqnarray}
Since $\om^iU_{\a}V_{\sbchi}\not\in Z(\S)$ for each $i\geq 0$, for some
$\om^rU_{\b}V_{\sbbeta}\in \S$ we must have $\bbeta(\a)\neq\bchi(\b)$.
This choice of $\om^rU_{\b}V_{\sbbeta}\in \S$ yields
$\bra{\psi}U_{\a}V_{\sbchi}\ket{\psi}=0$.
\end{enumerate}
\end{proof}

At this point it is useful to introduce a standard notation using which
it is convenient to describe quantum stabilizer codes. Let $\S$ be an
abelian subgroup of $\E$ with centralizer $Z(\S)$. The \emph{minimum
  distance} $d(\S)$ is defined to be the minimum of
\[
\{\wt(\a,\balpha)\mid \om^iU_{\a}V_{\sbalpha}\in Z(\S)\setminus\S\}.
\]

When $A$ is the additive abelian group of the finite field $\F_q$ we
define an $\brac{n,k,d}{q}$ quantum stabilizer code to be a
$q^k$-dimensional subspace $\C(\S)$ of $L^2(\F_q)^{\otimes^n}$, where
$\S$ is an abelian subgroup of $\E$ with $d(\S)\geq d$ and cardinality
$q^{n-k}$.

By Theorem~\ref{kl-3} it follows that an $\brac{n,k,d}{q}$ quantum
stabilizer code is a $\floor{(d-1)/2}$-error correcting quantum code.

\begin{remark}
  Let $\S$ be an abelian subgroup of $\E$ such that $\om^iI\not\in \S$
  for every $i\neq0$. This is equivalent to demanding that $\S$ is an
  abelian subgroup of $\E$ such that for any $\a\in A^n$ and
  $\bchi\in\hat{A}^n$ the operator $\om^iU_{\a}V_{\sbchi}$ can be in $\S$
  for at most one $i: 0\leq i\leq l-1$, Thus $\S$ has the form
\[
\S =\{p(\a,\bchi)U_{\a}V_{\bchi}\mid (\a,\bchi)\in S\}
\]
where $S\subset A^n\times\hat{A}^n$ is a subgroup satisfying
$\bchi(\a')=\bchi'(\a)$ for any $(\a,\bchi), (\a'\bchi')\in\S$ and $p$
is a function on $S$ with values in $\{\om^i\mid 0\leq i\leq l-1\}$.
\end{remark}

\section{Quantum stabilizer codes in the finite field setting}

In order to construct stabilizer quantum codes, we need to study
abelian subgroups $\S$ of $\E$ such that elements in $Z(\S)\setminus\S$ have
large weight. We choose $A$ to be a finite field $\F_q$, $q=p^r$ for
some prime $p$. In particular, the Hilbert space in which we seek
stabilizer codes is $L^2(\F_q)^{\otimes^n}$. Since $\F_q$ is an
abelian group under its addition operation with each nonzero element
of order $p$, it follows that every nontrivial character of $\F_q$ is
of order $p$. Choose a nontrivial character $\w\in\hat{F_q}$. Then
every other character $\om'\in\hat{\F_q}$ is of the form $\om_a$ where
$\om_a(x)=\w(ax)$ for all $x\in\F_q$. Likewise, every character in
$\hat{\F_q}^n$ is of the form $\om_{\a}$ where
$\om_{\a}(\x)=\w(\a\cdot\x)$ for all $\x\in\F^n_q$, where $\a\cdot\x$
is the inner product $\sum_i a_ix_i$, for $\a=(a_1,\ldots,a_n)^T$ and
$\x=(x_1,\ldots,x_n)^T$.

If we identify $\hat{\F_q}^n$ with $\F_q^n$, we can index the elements
of the error group $\E$ as $\om^iU_{\a}V_{\b}$, $0\leq i\leq p-1$, and
$\a, \b\in \F_q^n$, where $V_{\b}$ now stands for the operator
$V_{\sbchi}$ with $\bchi=\om_{\b}$. Thus, $\E$ is rewritten as
\[
\E = \{\om^iU_{\a}V_{\b}\mid 0\leq i\leq p-1, \a, \b\in \F_q^n\}.
\]
Notice that $\E$ is a finite group of cardinality $pq^{2n}$. The Weyl
commutation relations take the following form
\[
\w(\b\cdot\a)U_{\a}V_{\b}=V_{\b}U_{\a}\mbox{~~}\forall\mbox{~~} \a,\b\in\F_q^n.
\]

If $\S$ is a subgroup of $\E$ it is readily seen that $\S$ is abelian
if and only if for any two elements $\om^iU_{\a}V_{\b}$ and
$\om^jU_{\c}V_{\d}$ in $\S$ we have $\a\cdot\d=\b\cdot\c$. For
$(\a,\b)\in\F_q^n\times\F_q^n$, define $\wt(\a,\b)=\#\{i\mid
(a_i,b_i)\neq (0,0)\}$, where $\a=(a_1,\ldots,a_n)$ and
$\b=(b_1,\ldots,b_n)$. Let $S\subset \F_q^n\times \F_q^n$ be a
subgroup for which $\a\cdot\d=\b\cdot\c$ for all $(\a,\b), (\c,\d)\in
S$. Define
\[ 
S^{\perp_s}=\{(\a,\b)\in\F_q^n\times\F_q^n\mid
\a\cdot\d-\b\cdot\c=0\mbox{ for all }(\c,\d)\in\S\}. 
\]
Lemma~\ref{stab-1} and the Knill-Laflamme conditions can be restated
as follows.

\begin{lemma}\label{stab-2}
  Let $S\subset \F_q^n\times \F_q^n$ be a subgroup for which
  $\a\cdot\d=\b\cdot\c$ for all $(\a,\b), (\c,\d)\in S$. Suppose
  $\tilde{p}~:~S \rightarrow \{\om^i\mid 0 \leq i \leq p-1\}$ is a
  function such that $\S=\{\tilde{p}(\a,\b)U_{\a}V_{\b}\mid (\a,\b)\in
  S\}$ is an abelian subgroup of $\E$. Then $\C(\S)\subset
  L^2(\F_q)^{\otimes^n}$ is a quantum stabilizer code of dimension
  $q^n/{\#S}$. Furthermore, if $\wt(\a,\b)>2t$ for all nonzero
  elements $(\a,\b)\in S^{\perp_s}\setminus S$ then $\C(\S)$ is a
  $t$-error correcting quantum stabilizer code.
\end{lemma}

Thus the problem is to find subgroups $S$ of $\F_q^n\times\F_q^n$ such
that $\a\cdot\d=\b\cdot\c$ for all $(\a,\b), (\c,\d)\in S$ and
$\wt(\a,\b)$ is large for nonzero elements $(\a,\b)\in S^{\perp_s}\setminus S$.
The other problem is to ensure that we can build an abelian subgroup
$\S$ of $\E$ by picking a suitable $\tilde{p}~:~S \rightarrow
\{\om^i\mid 0 \leq i \leq p-1\}$ such that
$\S=\{\tilde{p}(\a,\b)U_{\a}V_{\b}\mid (\a,\b)\in S\}$. To this end,
we formulate an approach.

Let $\V$ be an $m$-dimensional vector space over $\F_q$ for a positive
integer $m$. Let $L:\V\rightarrow \F_q^n$ and $M:\V\rightarrow \F_q^n$
be two linear transformations. Thus, $L$ and $M$ can be written as
$n\times m$ matrices over $\F_q$. We restrict attention to abelian
subgroups of $\E$ that are of the form
\[
\{\tilde{p}(\v)U_{L\v}V_{M\v}\mid \v\in \V\}.
\]

Two elements $\tilde{p}(\v_1)U_{L\v_1}V_{M\v_1}$, 
$\tilde{p}(\v_2)U_{L\v_2}V_{M\v_2}$ commute precisely when

\begin{eqnarray}\label{eq-3}
\v_2^{T}L^{T}M\v_1 = \v_1^{T}L^{T}M\v_2 \mbox{~~}\forall\mbox{~~} \v_1, \v_2\in\V
\end{eqnarray}
and
\begin{eqnarray}\label{eq-4}
\frac{\tilde{p}(\v_1+\v_2)}{\tilde{p}(\v_1)\tilde{p}(\v_2)}=
{\w}(\v_2^{T}L^{T}M\v_1)=
{\w}(\v_1^{T}L^{T}M\v_2) \mbox{~~}\forall\mbox{~~} \v_1, \v_2\in\V.
\end{eqnarray}

Equation (\ref{eq-3}) will hold if we choose $L$ and $M$ such that
$M^{T}L=L^{T}M$ (i.e.\ $L$ and $M$ are such that $M^{T}L$ is
symmetric).

Writing $\tilde{p}(\v)={\w}(\tilde{q}(\v))$, for some function
$\tilde{q}:\V\rightarrow \F_q$, Equation (\ref{eq-4}) assumes the form
\[
\tilde{q}(\v_1+\v_2)-\tilde{q}(\v_1)-\tilde{q}(\v_2) = \v_2^{T}L^{T}M\v_1 
= \v_1^{T}L^{T}M\v_2 \mbox{~~}\forall\mbox{~~} \v_1, \v_2\in\V.
\]
For $p\neq 2$ we can choose $\tilde{q}$ to be the quadratic form
$\frac{1}{2}\v^{T}M^{T}L\v$. For $p=2$ the problem of recovering a
suitable quadratic form as a solution to the above equation is more
difficult. 

For the purpose of this article, we look for special solutions: we
demand that $L^{T}M$ be expressible as $D+D^{T}$ for some matrix $D$
over $\F_q$, which implies that $L^{T}M$ is a symmetric matrix with
diagonal entries as scalar multiples of 2. For example, we can choose
$D$ to be an upper diagonal matrix. Then $q(\v)=\v^{T}D\v$ is a
solution to Equation (\ref{eq-4}).  We summarize this below.

\begin{lemma}\label{lem-quad}
  Let $\V$ be a finite dimensional vector space, and $L:\V\rightarrow
  \F_q^n$ and $M:\V\rightarrow \F_q^n$ be two linear transformations
  such that $M^{T}L$ is symmetric and of the form $D+D^T$ for a linear
  map $D:\V\rightarrow \F_q^n$. Then 
\[
\S=\{\w(\v^TD\v)U_{L\v}V_{M\v}\mid \v\in\V\}
\]
is an abelian subgroup of the error group $\E$ on
$L^2(\F_q)^{\otimes^n}$.
\end{lemma}

An element $\om^iU_{\x}V_{\y}$ of $\E$ is in $Z(\S)$ if and only if
$\v^{T}M^{T}\x = \v^{T}L^{T}\y$ for all $\v\in\V$. Equivalently,
$\om^iU_{\x}V_{\y}\in Z(\S)$ if and only if $M^{T}\x=L^{T}\y$.

{From} the Knill-Laflamme conditions as stated in Theorem~\ref{kl-3},
$\C(\S)$ is a $t$-error correcting quantum code with $\S$ defined as
above if for any $(\x,\y)\in\F_q^n\times\F_q^n$, the condition
$M^{T}\x=L^{T}\y$ implies that either $\x=L\v$ and $\y=M\v$ for some
$\v\in\V$ or $\wt(\x,\y)>2t$.

If $\F_q$ has characteristic different {from} $2$ there is a partial
converse to Lemma~\ref{lem-quad}: Suppose $\C(\S)$ is some stabilizer
code in $L^2(\F_q)^{\otimes^n}$ where
$\S=\{\tilde{p}(\a,\b)U_{\a}V_{\b}\mid (\a,\b)\in\S\}$ for some
additive subgroup $S$ such that $\a\cdot\d=\b\cdot\c$ for all
$(\a,\b), (\c,\d)\in S$. Let $\#S=q^r$ and
$(\a_1\b_1),(\a_2,\b_2)\ldots,(\a_r\b_r)$ be an independent generating
set for $S$. Then $S_1=\{\a\in\F_q^n\mid \exists
\b\in\F_q^n~:~(\a,\b)\in S\}$ and $S_2=\{\b\in\F_q^n\mid \exists
\a\in\F_q^n~:~(\a,\b)\in S\}$ are linear subspaces of $\F_q^n$. Let
$\e_1,\e_2,\ldots,\e_r$ be the standard basis for $\F_q^r$. Define
$L:\F_q^r\rightarrow \F_q^n$ and $M:\F_q^r\rightarrow \F_q^n$ by
letting $L\e_i=\a_i$ and $M\e_i=\b_i$ for $i=1,2,\ldots,r$. Since ,
$\a\cdot\d=\b\cdot\c$ for all $(\a,\b), (\c,\d)\in S$, it follows that
$L^TM$ is symmetric. Suppose $\F_q$ is of characteristic $p\neq 2$.
For $(\a,\b)\in S$, let $\v\in\F_q^r$ be such that $L\v=\a$ and
$M\v=\b$ and define $p'(\v)={\frac{1}{2}}\v^TL^TM\v$. It is easy to
check that $p(\a,\b)=p'(\v)+\c\cdot\v$, for some $\c\in\F_q^r$. More
precisely, we have the following proposition.

\begin{proposition}
  Suppose $\F_q$ has characteristic different {from} $2$ and $\C(\S)$ is
  some stabilizer code in $L^2(\F_q)^{\otimes^n}$ of dimension
  $q^{n-r}$. Then there are linear transformations
  $L:\F_q^r\rightarrow \F_q^n$ and $M:\F_q^r\rightarrow \F_q^n$ such
  that $L^TM$ is symmetric and there is a $\c\in\F_q^n$ such that\\
  $\S=\{\w({\frac{1}{2}}\v^TL^TM\v+\c\cdot\v)U_{L\v}V_{M\v}\mid
  \v\in\F_q^r\}$.
\end{proposition}

We can derive the following proposition {from} Lemma~\ref{lem-quad}.

\begin{proposition}
  Let $L:\F_q^{n-1}\rightarrow\F_q^n$ be an injective linear map with
  range $C=\{\a\in\F_q^n\mid \sum_{i=1}^n a_i=0\}$ and $M=M'L$ for
  some symmetric linear map $M':\F_q^n\rightarrow \F_q^n$ of the form
  $M'=D+D^T$. Then
\begin{enumerate}
\item[(i)] $\S=\{{\w}(\a^TD\a)U_{\a}V_{M'\a}\mid \a\in C\}$
  is an abelian subgroup of $\E$. 
\item[(ii)] $\C(\S)$ is $t$-error correcting if for any
  $(\x,\y)\in\F_q^n\times\F_q^n$, the condition $\y-M'\x\in C^{\perp}$
  implies that either $\x\in C$ and $\y=M'\x$ or $\wt(\x,\y)>2t$.
\end{enumerate}
\end{proposition}

  
Let $\S=\{\w(\a^TD\a)U_{\a}V_{L\a}\mid \a\in C\}$, where
$L=D+D^{T}$, $L$ and $D$ are $n\times n$ matrices over $\F_q$, and $C$
is a subspace of $\F_q^n$. As already observed $\S$ is an abelian
subgroup of $\E$. Our next goal is to give an orthonormal basis for
$\C(\S)$. Notice that $\C(\S)$ is a $q^n/{\#C}$-dimensional subspace
of $L^2(\F_q)^{\otimes^n}$.  Since $C$ is an additive subgroup of
$\F_q^n$, it suggests that an orthonormal basis for $\C(\S)$ can be
indexed by the cosets of $C$ in $\F_q^n$. It suffices to describe unit
vectors $\ket{\psi_{C+\x}}\in L^2(\F_q)^{\otimes^n}$ that have
disjoint support in $\F_q^n$, and show that each $\ket{\psi_{C+\x}}$
is fixed by $\S$, where $\x$ runs over a set of distinct coset
representatives of $C$ in $\F_q^n$. Define
\begin{eqnarray}\label{eq-5}
\ket{\psi_{C+\x}} = 
\frac{1}{\sqrt{\#C}}\sum_{\a\in C}\w(\a^TD\a)\w(\a^TL\x)\ket{\a+\x}
\end{eqnarray}
for each coset $C+\x$ as $\x$ runs over a set of distinct coset
representatives of $C$ in $\F_q^n$. The vectors $\ket{\psi_{C+\x}}$
have unit norm, and as they have mutually disjoint supports, they form
an orthonormal set of $q^n/{\#C}$ vectors in $L^2(\F_q)^{\otimes^n}$.
It can be easily verified that $\S$ fixes each $\ket{\psi_{C+\x}}$.
We summarize or observations below.

\begin{proposition}
  Let $\S=\{{\w}(\a^TD\a)U_{\a}V_{L\a}\mid \a\in C\}$, where
  $L=D+D^{T}$, $L$ and $D$ are $n\times n$ matrices over $\F_q$, and
  $C$ is a subspace of $\F_q^n$. Then the collection of vectors
  $\{\ket{\psi_{C+\x}}\}$ defined as
\[
\ket{\psi_{C+\x}} = 
\frac{1}{\sqrt{\#C}}\sum_{\a\in C}\w(\a^TD\a)\w(\a^TL\x)\ket{\a+\x}
\]
for each coset $C+\x$ as $\x$ runs over a set of distinct coset
representatives of $C$ in $\F_q^n$, is an orthonormal basis for
$\C(\S)$. In particular, $\dim \C(\S) = q^{n - \dim C}$. 
\end{proposition}

\begin{remark}\label{punct}
  $\C(\S)$ is an $\brac{n,k,d}{q}$ quantum code if it has dimension
  $q^k$ and $d(\S)\geq d$. In line with classical coding theory we can
  define the \emph{rate} of an $\brac{n,k,d}{q}$ quantum code as $k/n$
  and \emph{relative distance} as $d/n$. It is clearly desirable to
  design quantum codes with large rates and relative distance. An
  $\brac{n,k,d}{q}$ quantum code $\C(\S)$ is a \emph{pure} code if the
  corresponding centralizer subgroup $Z(\S)$ has the property that
  $\wt(\a,\b)\geq d$ for each $\om^iU_{\a}V_{\b}\in Z(\S)$,
  $(\a,\b)\neq(0,0)$. (Notice that this is a stronger property than
  guaranteed by Theorem~\ref{kl-3}).  Given a pure quantum stabilizer
  code, the following simple method can be used for deriving new
  quantum codes.
  
  Suppose we have an $\brac{n,k,d}{q}$ pure quantum code, with a small
  $k$ and large $d$. {From} such a code we can construct an
  $\brac{n-1,k+1,d-1}{q}$ quantum code that is again pure, by the
  technique of puncturing $S$ to yield an additive subgroup $S'$ of
  $\F_q^{n-1}\times\F_q^{n-1}$, of size still $q^{n-k}$ and distance at
  least $d(\S)-1$.  The idea of punctured classical codes (see
  McWilliams and Sloane \cite{slone}) can be adapted to punctured pure
  quantum stabilizer codes following \cite{cald-shor} where it is shown
  for $q=2$. A repeated application of puncturing will give
  $\brac{n-k',k+k',d-k'}{q}$ codes for different choices of $k'$.
\end{remark}

\section{A class of stabilizer codes}\label{actual}

First choose and fix the following subspace $C$ of $\F_q^n$:
\[
C=\{(a_1,\ldots,a_n)^T\in\F_q^n\mid \sum_{i} a_i=0\}.
\]

The subspace $C$ is invariant under the cyclic shift permutation
$\sigma:i\mapsto (i+1)mod~n$. Thus,
$C^{\perp}=\{(a,\ldots,a)\in\F_q^n\mid a\in\F_q\}$ is also invariant
under $\sigma$. An $n\times n$ matrix $L$ over $\F_q$ is said to be
\emph{circulant} if for $i=2,\ldots,n$, the $\th{i}$ row of $L$ is
obtained by applying $\sigma^{i-1}$ to the first row.

Let $\S=\{\w(\a^TD\a)U_{\a}V_{L\a}\mid \a\in C\}$, where $L=D+D^{T}$
is an $n\times n$ matrix over $\F_q$ and $C$ is as chosen above. We
further specialize our construction by choosing $L$ to be an $n\times
n$ symmetric circulant matrix with entries {from} $\{0,1\}$ and with all
diagonal entries 0. Let $\e_1=(1,0,\ldots,0)\in\F_q^n$. For such an
$L$ observe that $\a^TL\e_1=\a^TD^T\e_1$ for $\a\in C$. Then the
orthonormal basis $\{\ket{\psi_{C+\x}}\}$ for $\C(\S)$ (as described
in Equation (\ref{eq-5})) can be written in the following form:

\begin{eqnarray}\label{eq-6}
\ket{\psi_{C+c\e_1}} = \frac{1}{\sqrt{\#C}}\sum_{\a\in C} 
\w((\a^T+c\e_1^T)D(\a+c\e_1))\ket{\a+c\e_1}, \mbox{~~~}\c\in\F_q.
\end{eqnarray}
In particular, for $q=2$ the above stabilizer code has a neat encoding
circuit that we describe in Figure~\ref{circ-fig} in the appendix.


As an example of stabilizer codes given by Equation~(\ref{eq-6}), we
now describe a $\brac{5,1,3}{q}$ quantum code for every finite field
$\F_q$. In particular, for $q=2$, the $\brac{5,1,3}{2}$ code is the
Laflamme code which was originally obtained by a computer search
\cite{Laf}. Let $L_5$ be the following symmetric circulant matrix in
$\F_q^{5\times 5}$.

\[
\left( \begin{array}{ccccc}
    0 & 0 & 1 & 1 & 0 \\
    0 & 0 & 0 & 1 & 1 \\
    1 & 0 & 0 & 0 & 1 \\
    1 & 1 & 0 & 0 & 0 \\
    0 & 1 & 1 & 0 & 0
       \end{array}   
\right)
\]
and $C=\{(a_1,\ldots,a_5)^T\in\F_q^5\mid \sum_{i} a_i=0\}$. It can be
checked that $S=\{(\a,L_5\a)\mid \a\in C\}$ is an additive subgroup of
$\F_q^5\times \F_q^5$ such that $d(\S)\geq 3$. Thus, by Theorem~\ref{kl-3},
$\C(\S)$ is a $\brac{5,1,3}{q}$ quantum code for every finite field
$\F_q$. The encoding circuit for the $\brac{5,1,3}{2}$ can be obtained
easily {from} the general encoding circuit already described for codes
given by Equation~(\ref{eq-6}). 

For a vector $\c\in\F_2^n$, let $\sigma\c\in\F_2^n$ denote the vector
obtained by a cyclic shift of $\c$. An $n\times n$ circulant matrix
with first column $\c\in\F_2^n$ can be conveiently written as
\[
\left( \begin{array}{cccc}
\c & \sigma\c &  \ldots & \sigma^{n-1}\c 
        \end{array}
\right)
\]
We give two more examples of quantum codes defined using circulant
matrices.

First, there is a $\brac{13,1,5}{2}$ quantum code defined by a
$13\times 13$ circulant matrix $L_{13}$ over $\F_2$, whose first
column is
\[
\c = (0, 0, 1,  1, 0, 0, 0, 0, 0, 0, 1, 1, 0)^T.
\]
As defined, $C=\{(a_1,\ldots,a_{13})^T\in\F_2^{13}\mid \sum_{i}
a_i=0\}$. It can be checked (with the help of a computer program) that
$S=\{(\a,L_{13}\a)\mid \a\in C\}$ is an additive subgroup of
$\F_2^{13}\times \F_q^{13}$ such that $d(\S)\geq 5$. Thus, by
Theorem~\ref{kl-3}, $\C(\S)$ is a pure $\brac{13,1,5}{2}$ quantum
code.

Similarly, there is a $\brac{21,1,7}{2}$ quantum code defined by a
$21\times 21$ circulant matrix $L_{21}$ over $\F_2$, whose first
column is
\[
\c =(0, 1, 1, 0, 1, 1, 1, 0, 0, 0, 0, 0, 0, 0, 0, 1, 1, 1, 0, 1, 1)^T
\]
As before, $C=\{(a_1,\ldots,a_{21})^T\in\F_2^{21}\mid \sum_{i}
a_i=0\}$. It can be checked using a computer program that
$S=\{(\a,L_{21}\a)\mid \a\in C\}$ is an additive subgroup of
$\F_2^{21}\times \F_q^{21}$ such that $d(\S)\geq 7$. Thus, by
Theorem~\ref{kl-3}, $\C(\S)$ is a pure $\brac{21,1,7}{2}$ quantum
code.

If $k=1$, it is interesting to note that for $n=5, 13$ and $21$, the
best achievable minimum distance \cite{cald-shor} is $d=3, 5$, and
$7$ respectively.

\section{Existence of good stabilizer codes}

Using a probabilistic argument we show that there is a number
$\alpha>0$ and a natural number $n_{\alpha}$ such that for each
$n>n_{\alpha}$ there exists a $\brac{n,1,\floor{\alpha n}}{2}$ pure quantum
stabilizer code. Now, as observed in Remark~\ref{punct}, given $\beta$
such that $0<\beta<\alpha$, by the method of punctured codes we can
obtain a family of $\brac{\floor{(1-\beta)n},\floor{\beta n},\floor{(\alpha-\beta)n}}{2}$
quantum codes for all $n>n_{\alpha}$. These are good quantum codes
with constant rate $\beta/(1-\beta)$ and constant relative distance
$(\alpha-\beta)/(1-\beta)$.

We first recall a particular form of the Chernoff bounds for bounding
the probability that a random variable deviates far {from} its
expectation. 

\begin{theorem}{\rm\cite[Theorem~4.2, page 70] {MR}}\label{chern}
  Let $X_1,X_2,\ldots,X_n$ be independent Bernoulli random variables
  such that for each $i$, $\Pr[X_i=1]=p$ and $\Pr[X_i=0]=1-p$, for
  $0<p<1$. Let $X=\sum_i X_i$ and let $\mu$ denote the expectation
  ${\bf E}[X]$.  Then for $0< \delta < 1$
\[
\Pr[X<(1-\delta)\mu] < e^{-\mu\delta^2/2}.
\]
\end{theorem}

Our existence proof for stabilizer codes will be guided by
Lemma~\ref{lem-quad}.

As before, we first choose and fix the following subspace $C$ of
$\F_q^n$:
\[
C=\{(a_1,\ldots,a_n)\in\F_q^n\mid \sum_{i} a_i=0\}.
\]

\begin{definition}\label{good}
  An $n\times n$ matrix $R$ over $\F_2$ is said to be
  $\alpha$-\emph{good} if the following conditions are true.
\begin{enumerate}
\item[(i)] The sum of every $\floor{\alpha n}$ columns of $R$ has weight at
  least $\alpha n$.
\item[(ii)] The sum of every $\floor{\alpha n}$ rows of $R$ has weight at
  least $\alpha n$.
\item[(iii)] The sum of every $\floor{\alpha n}$ columns of $R$ has weight at
  most $(1-\alpha)n$.
\item[(iv)] The sum of every $\floor{\alpha n}$ rows of $R$ has weight at
  most $(1-\alpha)n$.
\end{enumerate}
\end{definition}

As in classical coding theory \cite{slone}, given a vector
$\a=(a_1,a_2,\ldots,a_n)^T\in\F^n_q$ we denote $\#\{i\mid a_i\neq 0\}$
by $w(\a)$. The next proposition describes a way of constructing
stabilizer codes {from} good matrices.

\begin{theorem}\label{exists}
  For $0<\alpha<1$, suppose $R$ is an $n\times n$ $\alpha$-good matrix
  over $\F_2$.  Let $L$ be the following $2n\times 2n$ symmetric
  matrix over $\F_2$:
\[
\left( \begin{array}{cc}
       0 & R \\
       R^T & 0 
       \end{array}   
\right)
\]
If we write $L=D+D^T$, where $D$ is the upper triangular matrix with
zeros on the principal diagonal, and define the abelian subgroup $\S$
of $\E$ as $\S=\{\w(\a^TD\a)U_{\a}V_{L\a}\mid \a\in C\}$, then
$\C(\S)$ is a $\brac{2n,1,\floor{\alpha n}}{2}$ pure stabilizer code.
\end{theorem}

\begin{proof}
  {From} Lemma~\ref{lem-quad}, we know that $\C(\S)$ is a
  $\brac{2n,1,\floor{\alpha n}}{2}$ stabilizer code if for any
  $(\x,\y)\in\F_2^{2n}\times\F_2^{2n}$, the condition $\y-L\x\in
  C^{\perp}$ implies that either $\x\in C$ and $\y=L\x$ or
  $\wt(\x,\y)>\alpha n$. It is easy to check that the assumptions
  about $R$ in Definition~\ref{good}, in fact, guarantees a stronger
  property: for any nonzero vector $\x\in\F_2^{n}$ such that
  $w(\x)\leq\alpha n$, the assumptions (i) and (iii) imply that
  $\alpha n\leq w(R\x)\leq (1-\alpha)n$. Similarly, assumptions (ii)
  and (iv) imply that $\alpha n\leq w(R^T\x)\leq (1-\alpha)n$. Putting
  these together, it follows that $\alpha n\leq w(L\x)\leq
  (1-\alpha)n$ if $w(\x)\leq\alpha n$ for $\x\in\F_2^{2n}$.
  
  Since $C^{\perp}=\{(1,1,\ldots,1)^T,(0,0,\ldots,0)^T\}$, we can see
  that the above observation implies that $\wt(\x,\y)>\alpha n$ if
  $(0,0)\neq(\x,\y)\in\F_2^{2n}\times\F_2^{2n}$, and $\y-L\x\in
  C^{\perp}$. It follows that $\C(\S)$ is a $\brac{2n,1,\floor{\alpha
      n}}{2}$ pure stabilizer code. This completes the proof.
\end{proof}

We now show the existence of $n\times n$ matrices over $\F_2$ that
fulfill the conditions of Theorem~\ref{exists}.

\begin{lemma}\label{random}
  Let $R_{ij}$, $1\leq i,j\leq n$ be independent identically
  distributed random variables taking values in $\{0,1\}$ such that
  $\Pr[R_{ij}=1]=1/2$, $1\leq i,j\leq n$. Let $R$ be the uniformly
  distributed $n\times n$ random matrix over $\F_2$ whose $\th{ij}$
  entry is the random variable $R_{ij}$. There exist constants
  $\alpha>0$ and $n_{\alpha}>0$ such that
\[
\Pr[R\mbox{ is $\alpha$-good }]>0.
\]
\end{lemma}

\begin{proof}
  Let BAD denote the event that $R$ is not $\alpha$-good. Let
  $\r_1,\r_2,\ldots,\r_n$ be the rows of $R$ and
  $\c_1,\c_2,\ldots,\c_n$ be the columns of $R$. For any subset
  $S\subseteq\{1,2,\ldots,n\}$ with $1\leq \#S\leq \alpha n$, we
  define $E_S$, $D_S$, $A_S$, $B_S$ as the events $w(\bigoplus_{i\in
    S}\r_i)<\alpha n$, $w(\bigoplus_{i\in S}\r_i)>(1-\alpha)n$,
  $w(\bigoplus_{i\in S}\c_i)<\alpha n$, $w(\bigoplus_{i\in
    S}\c_i)>(1-\alpha) n$ respectively. Then BAD can be written as
  follows
\begin{eqnarray}\label{eq-7}
\BAD = \bigcup_{S\subset[n],1\leq \#S\leq \alpha n} 
A_S\cup B_S\cup D_S\cup E_S.
\end{eqnarray}
We analyze $A_S$ for a fixed $S$. Let $\bigoplus_{i\in
  S}\c_i=(x_1,x_2,\ldots,x_n)^T$. Since $R_{ij}$ $1\leq i,j\leq n$ are
all independent random variables taking values in $\F_2$,
$x_1,x_2,\ldots,x_n$ are $n$ independent uniformly distributed random
variables taking values in $\F_2$. We will use Chernoff bounds as
given in Theorem~\ref{chern} to analyze the random variable $\#\{i\mid
x_i=1, 1\leq i\leq n\}$. Let $X=\sum_{i=1}^nx_i$. Then ${\bf
  E}[X]=n/2$. Applying Theorem~\ref{chern} we get
\[
\Pr[A_S]=\Pr[X<\alpha n]\leq e^{-\frac{n(1-2\alpha)^2}{4}}.
\]

Notice that under $\F_2$ addition $1+x_1,1+x_2,\ldots,1+x_n$ are also
$n$ independent uniformly distributed random variables taking values
in $\F_2$. Thus, by Chernoff bounds we again obtain $\Pr[B_S]\leq
e^{-\frac{n(1-2\alpha)^2}{4}}$. Likewise, $\Pr[E_S]$ and $\Pr[D_S]$
are also bounded above by $e^{-\frac{n(1-2\alpha)^2}{4}}$. Putting
these together with the definition of BAD in Equation (\ref{eq-7}) we
get
\begin{eqnarray*}
\Pr[\BAD]\leq 4e^{-\frac{n(1-2\alpha)^2}{4}}
\cdot\sum_{i=1}^{\floor{\alpha n}}{{n}\choose{i}}
\leq 4e^{-\frac{n(1-2\alpha)^2}{4}} 2^{nH(\alpha)}
\end{eqnarray*}
where $H(\alpha)=-\alpha (\log\alpha)-(1-\alpha)\log(1-\alpha)$. To
ensure that $\Pr[\BAD]<1$, it suffices to pick $\alpha<1/4$ such that
$H(\alpha)<(\log e)3/8-2/n$, which can be done by choosing $n$ larger
than some constant $n_{\alpha}$ and $\alpha>0$ sufficiently small.
\end{proof}

{From} Theorem~\ref{exists}, Lemma~\ref{random}, and Remark~\ref{punct}
we can immediately deduce the following.

\begin{corollary}
  There are constants $\alpha>0$ and $n_{\alpha}>0$ such that for each
  $n>n_{\alpha}$ there is a $\brac{n,1,\floor{\alpha n}}{2}$ pure quantum
  stabilizer code. Furthermore, for any $\beta$ such that
  $0<\beta<\alpha$, and $n>n_{\alpha}$ there is a
  $\brac{\floor{(1-\beta)n},\floor{\beta
      n},\floor{(\alpha-\beta)n}}{2}$ pure quantum stabilizer code.
\end{corollary}

\begin{remark}
  The above existence argument can be easily extended to stabilizer
  codes over any finite field. More precisely, for $\F_q$ there are
  constants $\alpha>0$ and $n_{\alpha}>0$ such that for each
  $n>n_{\alpha}$ there is a $\brac{n,1,\floor{\alpha n}}{q}$ pure quantum
  stabilizer code.  Also, given a $\beta$ such that $0<\beta<\alpha$,
  and $n>n_{\alpha}$ there is a $\brac{\floor{(1-\beta)n},\floor{\beta
      n},\floor{(\alpha-\beta)n}}{q}$ pure quantum code.
\end{remark}

\noindent{\bf Acknowledgment}~~We thank Piyush P Kurur for his help
with the computer search for the codes given in Section~\ref{actual}.

\newpage

\begin{picture}(300,750)(90,17)%
\includegraphics*{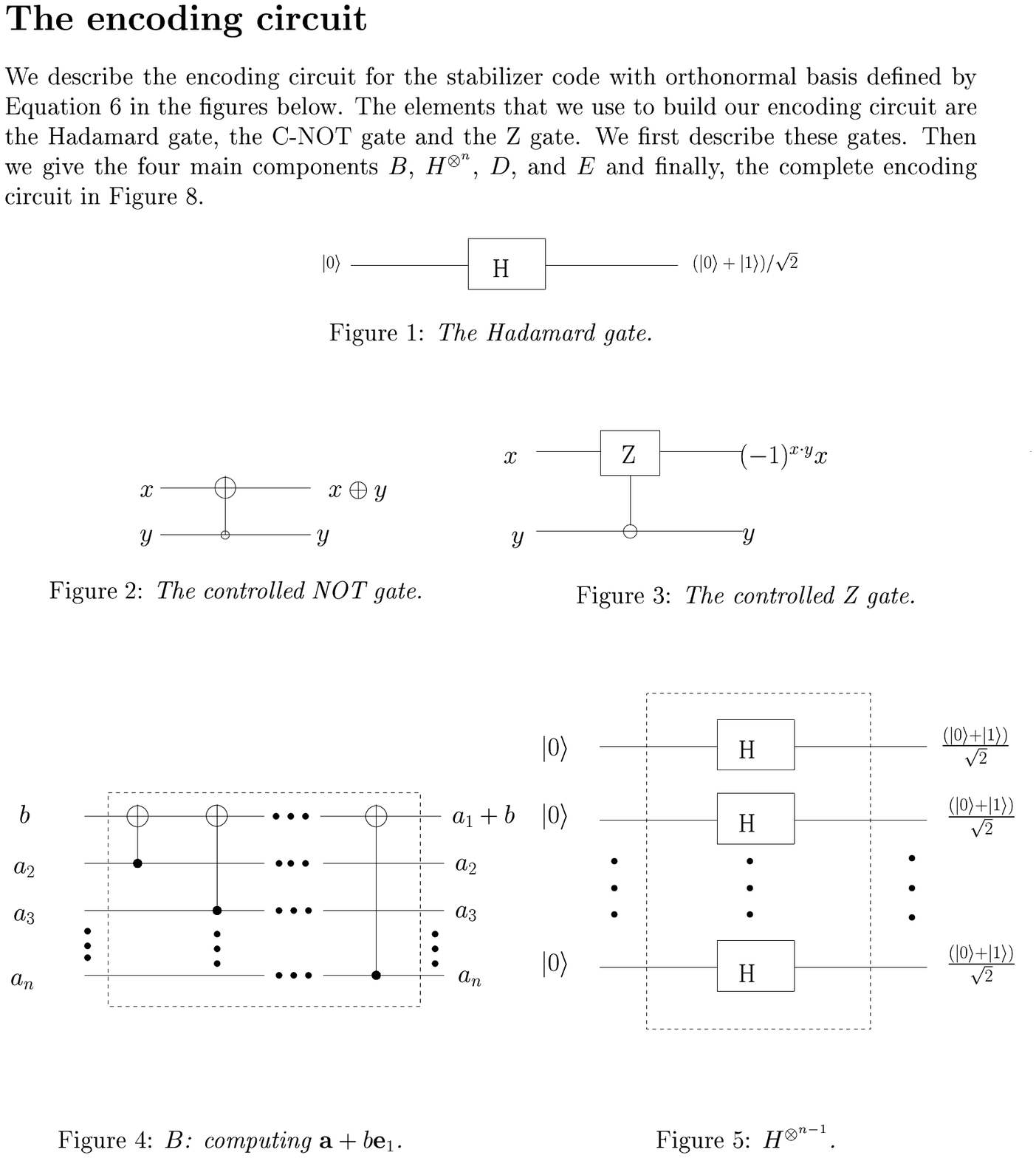}
\end{picture}

\newpage

\begin{picture}(300,750)(90,17)%
\includegraphics*{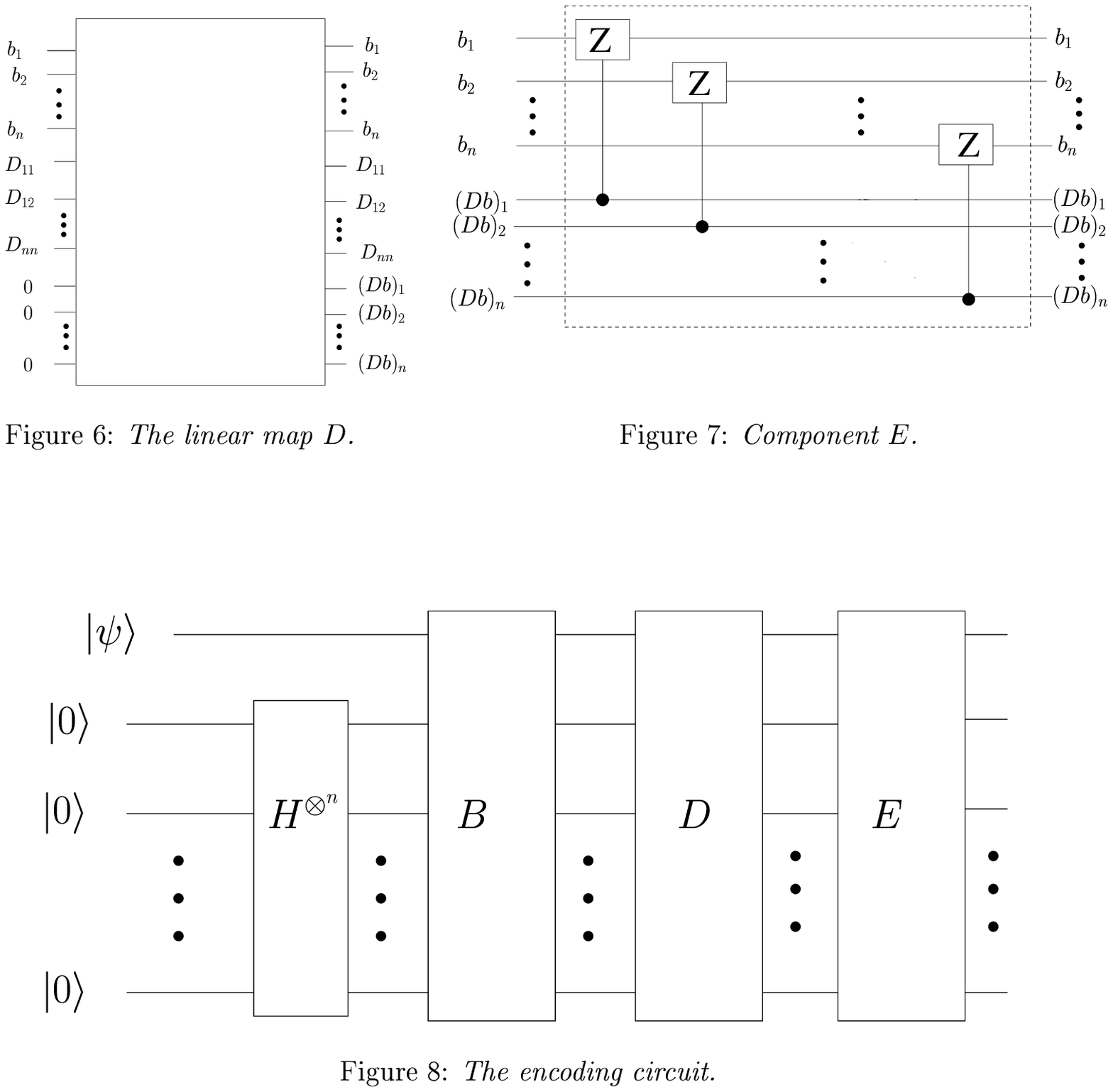}
\end{picture}

\end{document}